\documentclass[conference]{IEEEtran}
\IEEEoverridecommandlockouts
\usepackage{cite}
\usepackage{hyperref}
\usepackage{amsmath,amssymb,amsfonts}
\usepackage{algorithmic}
\usepackage{graphicx}
\usepackage{textcomp}
\usepackage{xcolor}
\usepackage{makecell}
\usepackage{siunitx}
\usepackage{booktabs}
\usepackage{caption}
\usepackage{tabularx}

\usepackage{soul}
\usepackage{authblk}
\usepackage{subcaption}
\usepackage{stfloats}
\def\BibTeX{{\rm B\kern-.05em{\sc i\kern-.025em b}\kern-.08em
    T\kern-.1667em\lower.7ex\hbox{E}\kern-.125emX}}
\begin{document}

\title{Traffic Chunk Sizing vs. Optical Switching Speed in Future All-Optical Satellite Networks}

\author{Sleman Mouammar}
\author{Thomas Röthig} %\thanks{A.A@university.edu}}
\author{Soheil Hosseini}
\author{Ítalo Brasileiro}
\author{Admela Jukan}
\affil{Technische Universität Braunschweig, Germany}

\maketitle

\begin{abstract}
In a transparent all-optically switched satellite networks, various switching paradigms can be considered, including packet, burst, or circuit. To determine the switching fabric design, the traffic assembly and path computations at the ground stations play a key role. Traffic can be buffered and assembled in chunks at the ground stations and forwarded over the pre-computed optical paths in space, following the principles known from terrestrial networks, like optical burst switching or fast circuit switching. Regardless of the paradigm chosen, the switching fabric is required to meet the latency requirements. This paper studies the traffic chunk sizing in future all-optical satellite networks as a function of the latency performance of optical switching fabrics. We consider various optical switching technologies, including MEMS- and integrated photonic-based solutions, their switching speed, power consumption, and insertion losses. Simulation results indicate that ns-switching speeds (e.g. SOA-based) enable larger chunk sizes of up to 500MB, than ms-order switches (e.g., MEMS) which enable up to 100 MB, when constrained by a typical maximum end-to-end latency of 60 ms.

\end{abstract}


\section{Introduction}  \label{sec:intro}

Modern satellite constellations, such as Starlink and Telesat, reportedly employ laser inter-satellite links (LISLs) together with packet-based electronically switched traffic \cite{Ma:23}. As transmission rates scale to hundreds of Gb/s, all-optical switching is expected to emerge enabling high-capacity all-optical networking in space. Akin to what is well-established in terrestrial networks, optical satellite networks can deploy either optical burst switching (OBS), optical packet switching (OPS), or optical circuit switching (OCS)  \cite{li2025review}. While different switching  paradigms result in different networking performance, the switching fabric in an optical satellite node plays a key role in the end-to-end latency performance. Parameters such as the number of orbital planes and satellites, constellation layout, and orbital altitude also contribute to end-to-end latency. 

%The fundamental challenge is therefore to enable effective transmission over all-optical paths in modern constellations while preserving low end-to-end (e2e) latency, which is one of the salient advantages of satellite networks.

%Fig. \ref{fig:scenario} illustrates the latency challenge. Let us assume that the traffic from an Internet service provider (ISP) needs to be assembled in traffic chunks at the ground stations to be sent over a satellite network. For every traffic chunk to be transmitted, in case that no immediate transmission is feasible, the traffic transmission needs to be scheduled over a path with an allocated wavelength. In a wavelength division multiplexing (WDM) system, one of the wavelengths can be allocated for data, % e.g., from $\lambda_1$ to $\lambda_4$, whereas a control channel is implemented out of band. 

\begin{figure}[t]
    \centering
    \includegraphics[width=1\linewidth]{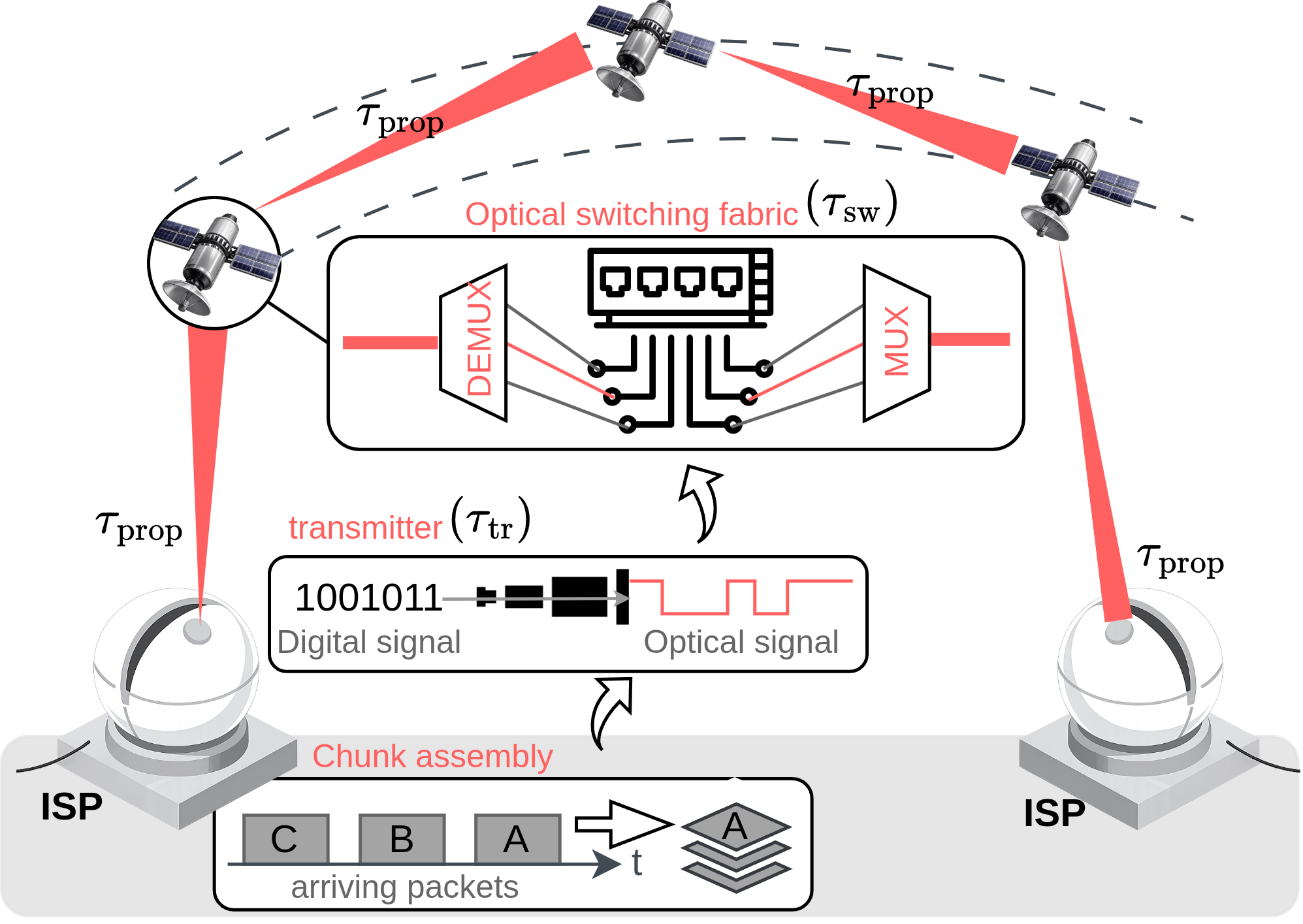}
    \caption{Traffic chunk sizing and onboard optical switching in next-generation all-optical satellite networks.}
    \label{fig:scenario}
\end{figure}

Fig. \ref{fig:scenario} illustrates the latency challenge. Let us assume that the traffic from an Internet service provider (ISP) needs to be assembled in traffic chunks at the ground stations to be sent over a satellite network. For every traffic chunk to be transmitted, in case that no immediate transmission is feasible, the traffic transmission needs to be scheduled over a path with an allocated wavelength. In a wavelength division multiplexing (WDM) system, one of the wavelengths can be allocated for data, whereas a control channel is implemented out of band. The end-to-end (e2e) latency is governed by three main delay components: propagation $\tau_{prop}$, transmission $\tau_{tr}$ and switching $\tau_{sw}$. The propagation delay depends on the speed of light in vacuum/atmosphere and the distance. Switching delay depends on the switch device technology, while the transmission delay, defined as the time when the transmission commences and ends,  depends on both the achievable data rate and the chunk size. To guarantee the end-to-end latency withing limits of what is acceptable, the transmission and switching delays need to be balanced, while maintaining high link utilization.
%Onboard the satellite, the optical switching fabric forwards the signal from input towards one of the outputs, based on the control information which contains the switch configuration instructions.
%\hl{this paragraph says NOTHING about the latency challnege? what is the narrative here which leads to the problem of maximum chuck size?? you need to describe the PROBLEM on HOW MAX CHUNJ SIZE IMPACTS THE LATENCY then continue with paragraph 3}

In this paper, we study the maximum size of the traffic chunk to be transmitted over the all-optical satellite network, for a given e2e delay constraint.  We model the optical satellite switching mechanism similar to OBS, whereby the transmission chunk is sent after a control message which ahead of the data transmission configures the switches along the path. We evaluate the impact of transmitting different data chunks in different scenarios of onboard optical switching devices, and analyze the performance in terms of chunk feasibility ratio and energy efficiency, subject to the e2e delay constraint.
%thus testing variable switching speeds, power consumption \hl{ the problem of power is not clear? do you minimize power?}, and network efficiency \hl{what is that>>>????}. 
%We model the switching mechanism similar to OBS, when adopting separated data and control packets, with the control packet being transmitted in a dedicated channel, followed by the data packet/data chunk. 
%\hl{differences to what??? you need to mention either the paper you are differnt from or something well known}  
%We evaluate the related switching parameters, ensuring that switching performance is aligned with traffic chunk sizing and e2e delay. \hl{this sentence is non-information. }
The paper considers three representative LEO constellations: Telesat, Amazon LEO Shell 1, and Starlink Phase 1. The results show that the maximum feasible data chunk for devices with ns-level switching delay is around 600MB in the evaluated constellations, whereas for ms-level switch, the maximum chunk size is 100 MB under 60 ms maximum e2e latency. 
%\hl{be specific and to the point}. %Small traffic chunks do not optimally utilize the optical spectrum, motivating research for new switching technology.

The rest of the paper is organized as follows. Section II provides related work. In Section III, we present a reference architecture. Section IV presents the simulation setup and performance evaluation. Finally, Section V  concludes the paper.

\section{Related Work}  \label{sec:realtedWork}
Although optical switching paradigms have been extensively investigated in terrestrial networks, their adoption in satellite systems remains relatively limited. To the best of our knowledge, none of the related work assess the impact of different data chunk sizes on e2e latency defined by a realistic threshold of maximum delay as observed in real-world satellite constellations. Recent work examines other important mechanisms. Wang et al. \cite{Wang-QoS-2019} introduce a quality of service (QoS)-adaptive burst assembly algorithm for space-based OBS networks, reducing assembly delay for high-priority traffic by up to 14.68\%. Similarly, \cite{Wang-OSBN-2022} proposes an optical switched satellite backbone network combining time-division multiplexing with OBS to enable QoS-aware resource allocation, achieving low latency and near-zero packet loss under specific system configurations.

%\hl{Neither of these works, however, relates the burst formation to hardware constraints and switching-related aspects. IN THE INTRO IT IS NOWHERE SAID THAT IT IS THE GOAL> INSTEAD THE TALK IS ABOUT LATENCY}
% Zhao et al. \cite{zhao:22} investigate the impact of burst size within a single data channel in an OBS-based satellite network with LISLs and RF feeder links, showing that OBS outperforms OCS in terms of loss and link utilization for burst durations of 100, 200, and 500 $\mu s$. However, their analysis is limited to single-channel scenario and does not consider multi-channel architectures or switching fabric constraints.

\begin{figure*}[hb]
    \centering    
    \includegraphics[width=.9\linewidth]{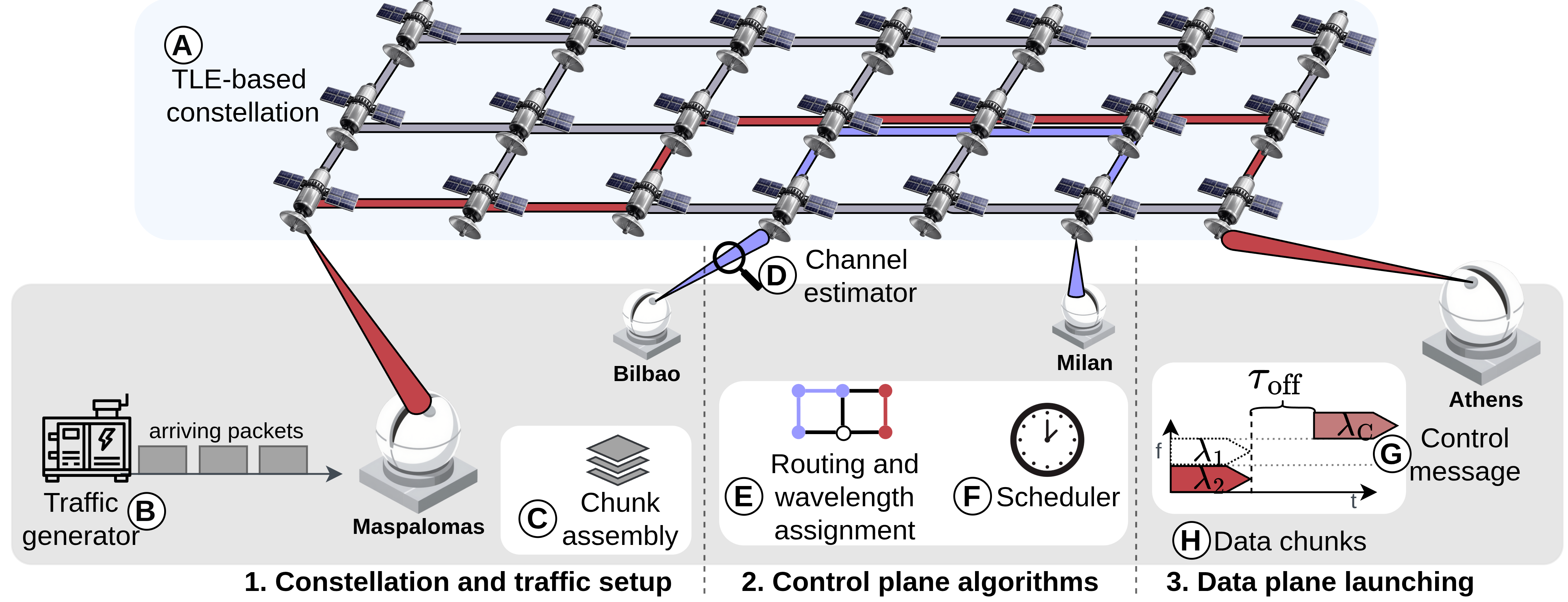}
    \caption{Reference architecture} %\hl{this figure is now redundant with fig 1 needs to be redrawn to include parts which are here...for instance expand on TLE channel estimator etc...as opposed to repeating figure 1}.}%\hl{rename control packet into control message.}}
    \label{fig:architecture}
\end{figure*}

The work in \cite{Li-optical-switching-2026} designs and evaluates the impact of SOA-based ultrafast OPS switches that operates with ns-level switching and channel capacity of 100 Gb/s, under a multi-granular dynamic traffic scenario. The work evaluates performance in terms of throughput, packet loss and contention, and propose a new scheduling algorithm to support OPS and TDM in satellite constellations. The latency evaluation is limited to the delay added by the specific OPS switching device, which we also consider but as part of the e2e latency. As the switching delay is one of the key components of the e2e delay in satellite networks, we evaluate the e2e latency performance, not only of SOA optical switches but also of other switching technology, to gain insights of the impact of optical switching technology and hardware choice on satellites.
%In parallel, advances in laser communication and onboard optical switching eliminate OEO conversion, thereby reducing latency, power consumption, and congestion \cite{Li-optical-switching-2026}.  However, feasible engineering and realization depend critically on the underlying switching fabric. \hl{the previsou two sentences do not tell me how is this related to our paper and what exectely is meant by the second sentence?! be specific}

Optical switching technologies such as Micro-Electro-Mechanical Systems (MEMS) and integrated photonic platforms offer distinct trade-offs in switching speed, power consumption, insertion loss, and system complexity \cite{Cheng-switching-2018}. While modern optical switches can achieve sub-microsecond to nanosecond reconfiguration times with reduced power consumption, prior work focused on device-level performance, without evaluating the impact added to the system in terms of e2e delay, which in satellite network plays a critical role.
%\hl{how do you define system level implications, and how does that related to the goal of the paper as stated i nthe intro? rephrase.}
Experimental efforts have started to address this gap, such as the work in \cite{zhai2021design} which validates an OBS-based satellite payload using an arrayed waveguide grating router, demonstrating transparent optical switching with wavelength conversion capabilities. However, their system is limited by a reconfiguration interval of 11.9 $\mu s$ and relies on wavelength conversion, which creates "electronic bottlenecks" in the network and leads to high power consumption and latency \cite{Li-optical-switching-2026}.
%which may not be compatible with strict Size, Weight, and Power (SWaP) constraints.\hl{this last statement is arbitrary and needs to be backed by a reference.} % \cite{Feyisa-soa-2022, Aida-switch-2022}

The overall e2e latency in all-optical satellite constellations also depends on the adopted traffic switching and forwarding paradigm, which in turn impacts the application performance. While OPS and OBS technologies are not technologically mature due to the need for optical buffering, OCS and its variations, like fast circuit switching, enable transparent switching once the path is reserved \cite{li2025review, zhao2020feasibility}. In this paper,  we consider resource allocation mechanisms similar to either fast OCS, or OBS with path reservation (no contention resolution at nodes) in which a control message pre-reserve a channel for a given period to a specific chunk transmission or circuit duration.

\section{Reference architecture} \label{sec:scenario}

Fig. \ref{fig:architecture} illustrates the reference architecture, which contains three important subsystems: \emph{Constellation and traffic setup}, \emph{Control plane algorithms}, including routing, resource reservation, and scheduling; and \emph{Data plane launching}, including the transmission of data chuck after the control message.

\subsection{Constellations and traffic setup}\label{sec:A}

The first setup component to consider is the constellation configuration (A). The Constellation Three-Line Element (TLE) file is commonly adopted by applications and datasets to represent satellite nodes. It typically represents mean orbital elements derived from tracking data and is formatted for compatibility with propagation models. Although satellites move, the topology at any time window can be captured and fixed in a TLE file or a snapshot. The TLE file describes topological aspects of satellite nodes such as altitude, density, epoch, satellite ids, and others that serve as input for simulations, propagation models and constellation snapshots.

In Fig. \ref{fig:architecture}, a traffic generator (B) is placed in every of the four optical ground stations (OGSs), which here exemplify traffic sources that generate data chunks, and are added to the OGS processing queue. In this work, we consider that every OGS is a possible source or destination of a traffic chunk, and traffic generators are placed in all ground nodes. 
Each data chunk have an associated control message, which provides information such as the destination OGS and other transmission requirements.
This same control message is updated before transmission with added information on scheduling, path, and spectral resources to allocate.
%When arriving at the OGS, the data chunks with fixed size are directed to high-capacity buffers and grouped according to the specified destination node. These buffers assemble the data into traffic chunks with fixed size (C), and the assembly is limited by either a time limit or a maximum chunk size. 

Specific constellation layouts, such as the Grid+ are considered to have fixed and stable inter-satellite links. But the uplink and downlink connections are still dynamically established based on line-of-sight and OGS placement, thus are checked periodically in order to verify availability.

\subsection{Control plane algorithms} \label{subsec:control}

As illustrated in Fig. \ref{fig:architecture}, a channel estimator (D) uses information such as link distance, SNR from up-, down, and LISLs, and associated interferences to estimate the feasible transmission bitrate. Based on this information, a path computation component (E) at the OGS runs routing and wavelength reservation algorithms to select appropriate lambdas (e.g., from $\lambda_1$ to $\lambda_2$) in the optical up- and downlink for chunk transmission. When route and spectral resources are decided, a scheduler (F) runs an algorithmic solution to determine the best time for launching that chunk. The channel capacity (D) for each of the LISLs, as well as the up- and downlink, is estimated based on the following channel model. The up- and downlink data rates are expressed as a function of the signal-to-noise ratio (SNR) and the bandwidth $B$ as:
\begin{equation}\label{eq:cap}
    C = B \log_2(1 + \mathrm{SNR})
\end{equation}
The SNR is determined by the received power $P_R$ and the system noise power $N_0$, and is given by
\begin{equation}\label{eq:SNR}
    \mathrm{SNR} = \frac{P_R}{N_0}.
\end{equation}
The received power is obtained from the link-budget formulation, based on \cite{Spirito:25, cantore2024link}:
\begin{equation}
P_R =
\frac{P_T}{N_{\mathrm{ch}}}
G_T \eta_T G_R \eta_R
L_{\mathrm{FSO}} L_{\mathrm{atm}} L_{\mathrm{fc}} L_{\mathrm{p}},
\label{eq:received_power}
\end{equation}

where $G_T$ and $G_R$ denote the transmitter and receiver gains, respectively, while $\eta_T$ and $\eta_R$ represent their corresponding optical efficiencies. Furthermore, $L_{\mathrm{FSO}}$ denotes the free-space optical path loss, $L_{\mathrm{atm}}$ accounts for atmospheric effects, including scintillation-induced beam wander and turbulence, and $L_{\mathrm{p}}$ represents the pointing-error loss. The term $L_{\mathrm{fc}}$ denotes the additional coupling or channel-related loss considered in the link budget.
Each loss component is briefly described below. During propagation from the OGS to the satellite, the signal is attenuated by several factors, starting with the free-space path loss (FSPL), which depends on the slant range $R(\vartheta_{\mathrm{el}})$, elevation angle $\vartheta_{\mathrm{el}}$, and wavelength $\lambda$:

\begin{equation}
L_{\mathrm{FSO}} =
\left(
\frac{\lambda}{4\pi R(\vartheta_{\mathrm{el}})}
\right)^2
\label{eq:fso_path_loss}
\end{equation}

The signal also experiences atmospheric attenuation caused by scattering from particles along the propagation path. Using Kim's model, this loss depends on the geometrical scattering attenuation coefficient $\alpha_{\mathrm{scatt}}(V,\lambda)$, where $V$ is the meteorological visibility expressed in km, and the Mie scattering defined based on the extinction ratio $\rho$ and elevation angle $\vartheta_{\mathrm{el}}$. It is given by \cite{Spirito:25, Erdogan:21}:

\begin{equation}
L_{\mathrm{atm, att}} = \alpha_{\mathrm{scatt}}(V,\lambda) R(\vartheta_{\mathrm{el}}) +  \exp(-\frac{\rho}{\sin(\vartheta_{\mathrm{el}})})
\end{equation}
Finally, in the uplink, atmospheric turbulence can cause scintillation-induced beam wander, leading to misalignment loss \cite{cantore2024link}. This loss is characterized using the Gamma-Gamma distribution, where $\alpha$ and $\beta$ denote the small- and large-scale turbulence parameters, respectively \cite{Spirito:25}. The received optical intensity is denoted by $I$, while $\Gamma(\cdot)$ and $K(\cdot)$ represent the Gamma function and the modified Bessel function, respectively \cite{Spirito:25}.
\begin{equation}
    f_{GG}(I) = \frac{2(\alpha\beta)^{(\alpha+\beta)/2}}{\Gamma(\alpha)\Gamma(\beta)}I^{\frac{\alpha+\beta}{2}-1}K_{\alpha-\beta}(2\sqrt{\alpha\beta I})
\end{equation}
For the uplink, the atmospheric loss is modeled as the combination of atmospheric attenuation and scintillation-induced beam-wander loss, i.e.,
\begin{equation}
    L_{\mathrm{atm}} = L_{\mathrm{atm,att}} + L_{\mathrm{sc}}
\end{equation}
In contrast, for the downlink, only atmospheric attenuation is considered \cite{cantore2024link}, yielding
\begin{equation}
    L_{\mathrm{atm}} = L_{\mathrm{atm,att}}
\end{equation}
Pointing error losses for  LISLs and up- and downlink are calculated based on the Irradiance displacement model from~\cite{Shang:25}:

\begin{equation}
f_{h_{PE}}(h)
= \gamma^2
 A_0^{- \gamma^2}
  h^{\gamma^2-1},
\quad 0<h\le A_0.
\label{eq:pointing_loss}
\end{equation}

with $A_0 = [\mathrm{erf}(v)]^2$ and $v = \frac{\sqrt{\pi}w_d}{\sqrt{2}w_z}$, $\gamma = \frac{w_{\mathrm{eq}}}{2\sigma_p}$ and $w_{\mathrm{z_{eq}}}^2=w_z^2\frac{\sqrt{\pi}\mathrm{erfc}(v)}{2v\exp(-v^2)}$.
In which $\omega_d$ is the receiving aperture radius, $\omega_{\mathrm{zeq}}$ is the equivalent beam width, and $\sigma_p$ is the jitter standard deviation. Using the pointing loss generated from the channel model described by the distribution in Eq.~\ref{eq:pointing_loss}, the received power for the LISL is given by:

\begin{equation} \label{eq:received_power_lisl}
    P_R = \frac{P_T}{N_{ch}} G_T G_R \eta_T\eta_RL_{FSO}L_{fc}L_{p}
\end{equation}

Since LISLs establish direct optical communication between satellites, the transmitted signal does not propagate through the Earth's atmosphere. Therefore, atmospheric attenuation is not considered in the LISL when calculating capacity. Thus, the capacity can be obtained from Eq.~\ref{eq:cap}, using the SNR defined in Eq.~\ref{eq:SNR}. The received power varies for feeder links and LISL based on Eqs. \ref{eq:received_power} and \ref{eq:received_power_lisl} respectively. % When selecting the path for a given chunk, the channel estimator calculates the achievable data rate for each link on a path and the bottleneck link is identified. 

Fig.~\ref{fig:architecture} depicts a dedicated and out-of-band control wavelength $\lambda_C$ for control messages (G). The wavelengths $\lambda_{1}$ and $\lambda_{2}$ are used in the data plane.
%; such separation is common in terrestrial optical networks. 
%It should be noted that a control channel is out-of-band (either RF or dedicated optical), and dedicated to control messages (G). 
Control messages are transmitted ahead of the corresponding data traffic (after offset time $\tau_{off}$ has passed, represented by the chunks in (H)) and contain information on scheduling and resources to reserve. Control messages undergo optical-to-electrical (OE) conversion $\tau_{OE}$, electrical processing, and electrical-to-optical (EO) conversion $\tau_{EO}$ at each satellite node, and setup the optical switch fabric for a proper switching of its corresponding data chunk. Since the estimated capacity limits the data rate possible over an e2e path, it is utilized as lower bound to calculate the transmission delay based on the chunk size to be transmitted.

\subsection{Data plane launching}

When the data is ready to launch, a control message (G) is sent over $\lambda_C$ and reserves the computed resources obtained from (E). Afterwards, the transmission initiates (H), for which we then measure the resulting e2e latency, i.e., the time between data transmission and data reception at the destination. The e2e latency considers three main delay components: propagation, transmission, and switching delays. Since the processing delay of the control message is considered constant it is added as offset delay to the e2e delay.

The transmission delay $\tau_{tr}$ corresponds to the time required to inject a data payload onto the physical medium. Thus, larger payloads or lower-data-rate links contribute to this delay. %, with the bottleneck link limiting the path’s effective throughput. 

The switching delay $\tau_{sw}$ represents the reconfiguration time of the optical switching fabric. This parameter varies significantly with underlying hardware, ranging from milliseconds for MEMS to nanoseconds for SOAs. The subsequent analysis in this paper quantifies the impact of multiple device-dependent switching delays on overall system performance. The propagation delay $\tau_{prop}$ is determined by the speed of light across uplink, downlink, and LISLs, and scales with orbital altitude, constellation density, and inter-satellite distances. Finally, the processing delay $\tau_{proc}$ is defined as the time required to process and execute control-plane signaling.
%, which is typically transmitted in advance to pre-configure the optical switching fabric for incoming data chunks.

The e2e delay performance of all-optical satellite networks depends greatly on the optical switching fabric deployed and the chunk size transmitted, and presents a fundamental design trade-off. Whereas larger data chunks occupy network resources for longer times and can increase the network load, smaller chunks may lead to poor link utilization with non-negligible switching time, and e2e latency becomes dominated by switching times rather than data transmission time. Thus, the chunk size plays a key role in determining the trade-off between switching speed and e2e latency. To study this trade-off for various switching fabrics, we analyze the chunk feasibility ratio metric against e2e latency. The chunk feasibility ratio (CFR) is defined as the percentage of  chunks that can be successfully allocated and are not dropped due to link unavailability, out of the total number of generated chunks,~i.e.,
\begin{equation}\label{eq:BR}
    \mathrm{CFR} = (1 - \frac{N_{\mathrm{Dropped}}}{N_{\mathrm{Total}}}) * 100\%
\end{equation}
As previously mentioned, e2e latency ($L$) is the sum of the following delay components on an optical path, i.e., transmission delay $\tau_{tr}$ at the bottleneck link, propagation delay $\tau_{\mathrm{prop}}$, and onboard satellite delays, including processing and optical switching, i.e., 
\begin{equation}\label{eq:L}
    L = \tau_{tr} + \sum_{\mathrm{Hops}} \tau_{\mathrm{prop}} + \sum_{\mathrm{Sats}} \left( \tau_{\mathrm{proc}} + \tau_{\mathrm{switch}} \right)
\end{equation}

The presented CFR in Eq. \ref{eq:BR} and e2e latency (Eq. \ref{eq:L}) metrics allow analyzing how the different chunk sizes affect the resource availability and allocation, as bigger data chunks imply higher transmission delays, longer channel reservation, and increased competition for free resources.

\section{Numerical results} \label{sec:performance}

\begin{figure*}[ht]
    \centering
    \begin{subfigure}[b]{0.32\textwidth}
        \includegraphics[width=\textwidth]{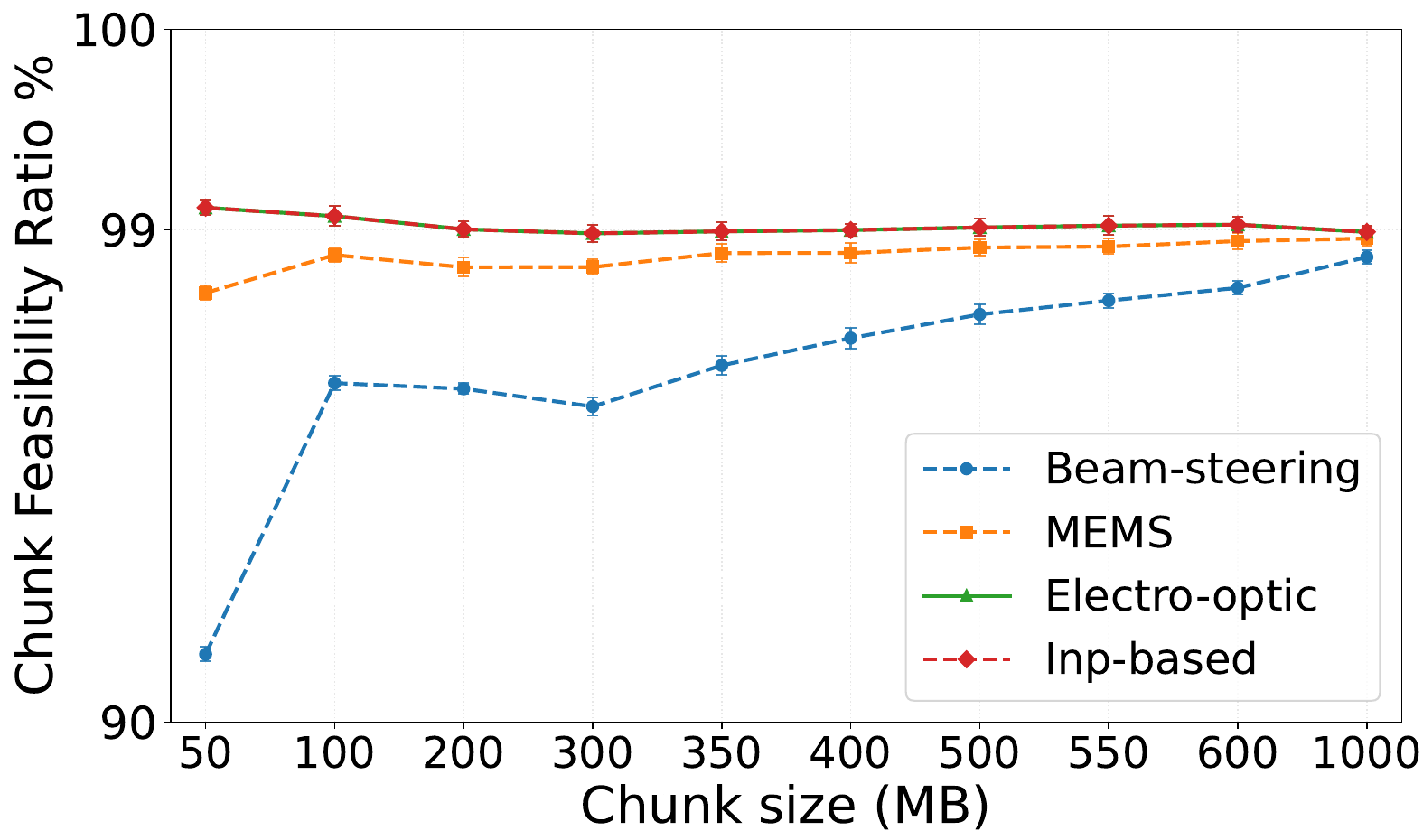}
        \caption{Telesat}
        \label{fig:fig1.a}
    \end{subfigure}
    \begin{subfigure}[b]{0.32\textwidth}
        \includegraphics[width=\textwidth]{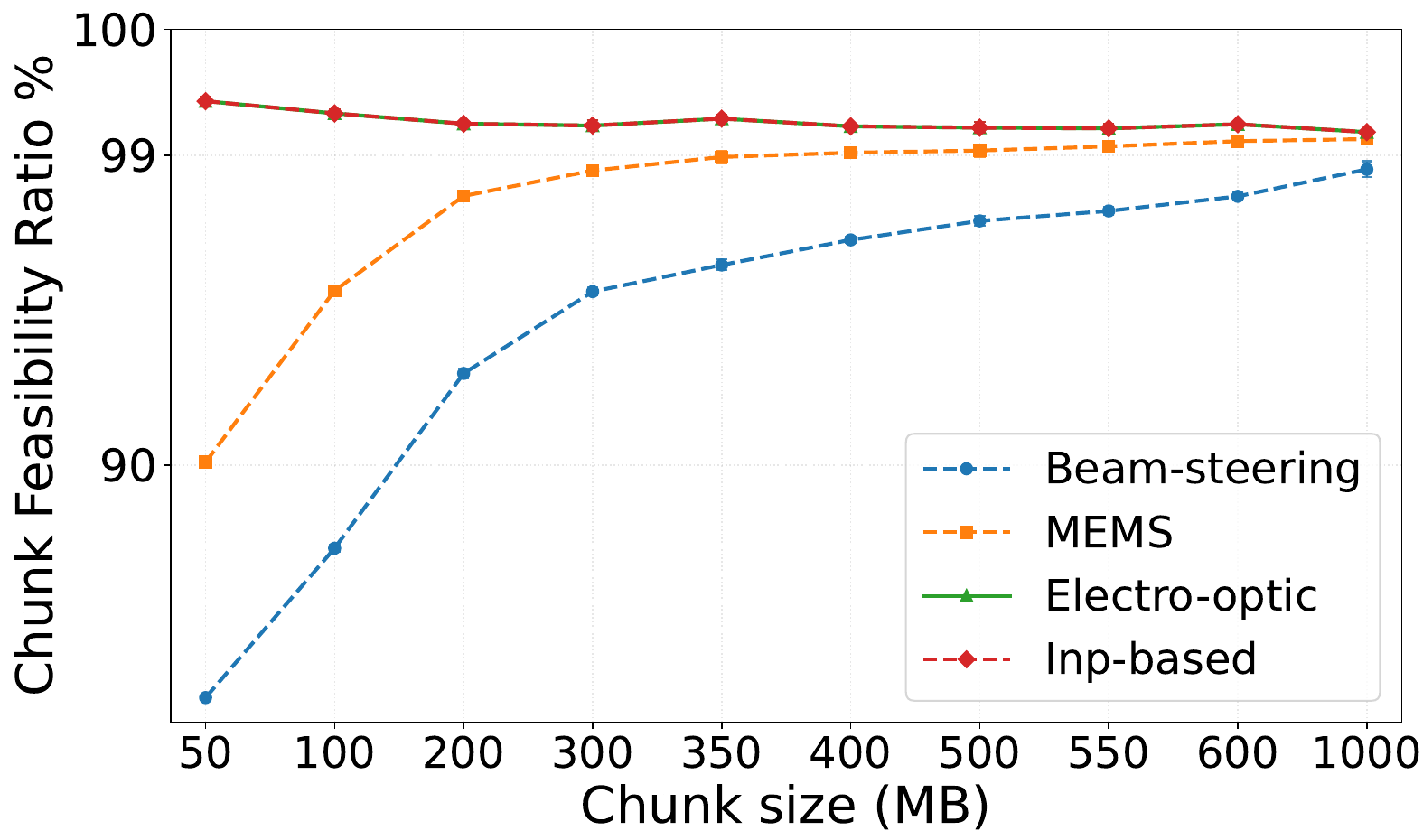}
        \caption{Amazon Leo shell 1}
        \label{fig:fig1.b}
    \end{subfigure}
    \begin{subfigure}[b]{0.32\textwidth}
        \includegraphics[width=\textwidth]{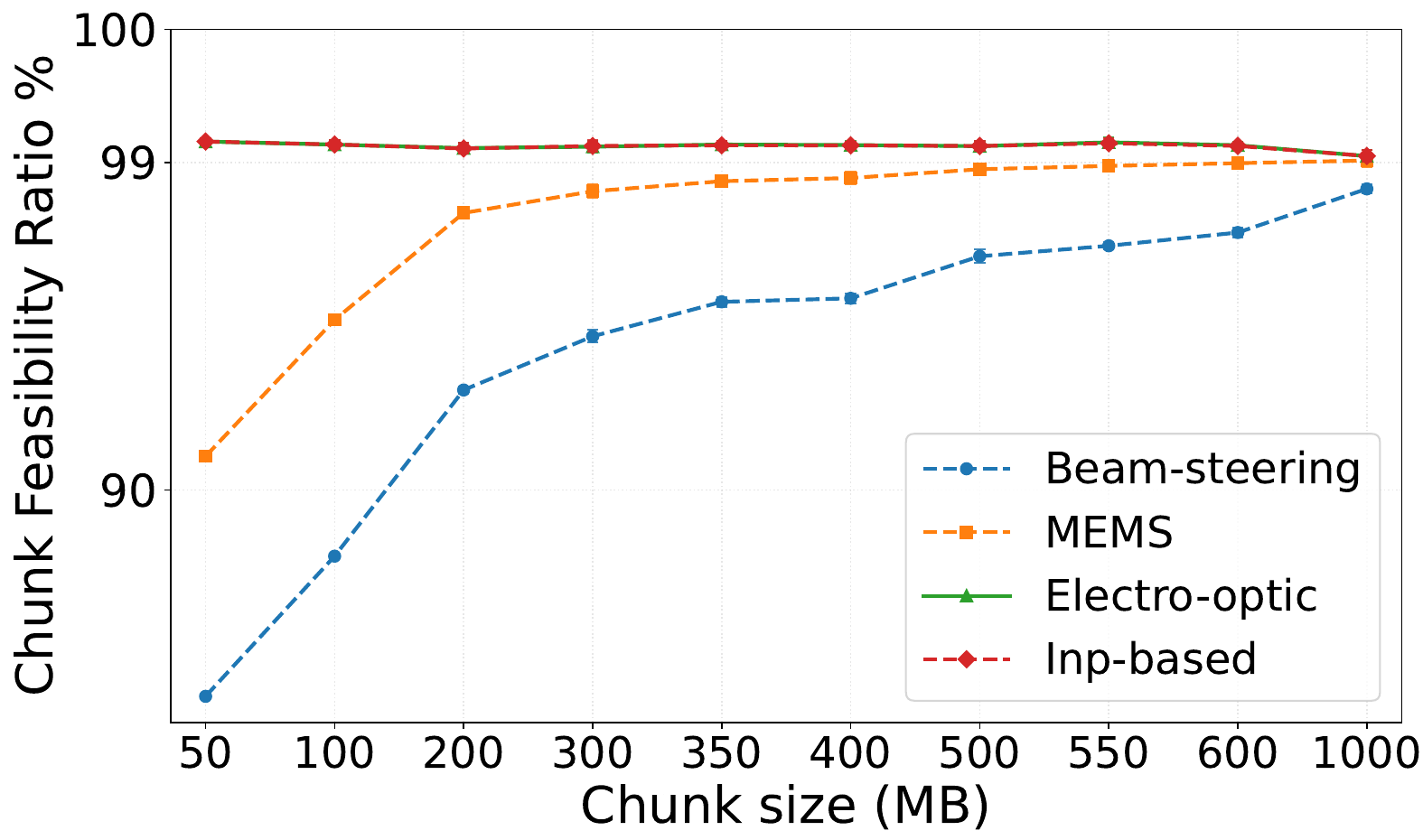}
        \caption{Starlink phase 1 }
        \label{fig:fig1.c}
    \end{subfigure}
    \caption{Chunk feasibility ratio for different chunk sizes vs.  switching fabrics for various constellations}
    \label{fig:fig1}
\end{figure*}

\begin{figure*}[ht]
\centering
\begin{subfigure}[b]{0.32\textwidth}
    \includegraphics[width=\textwidth]{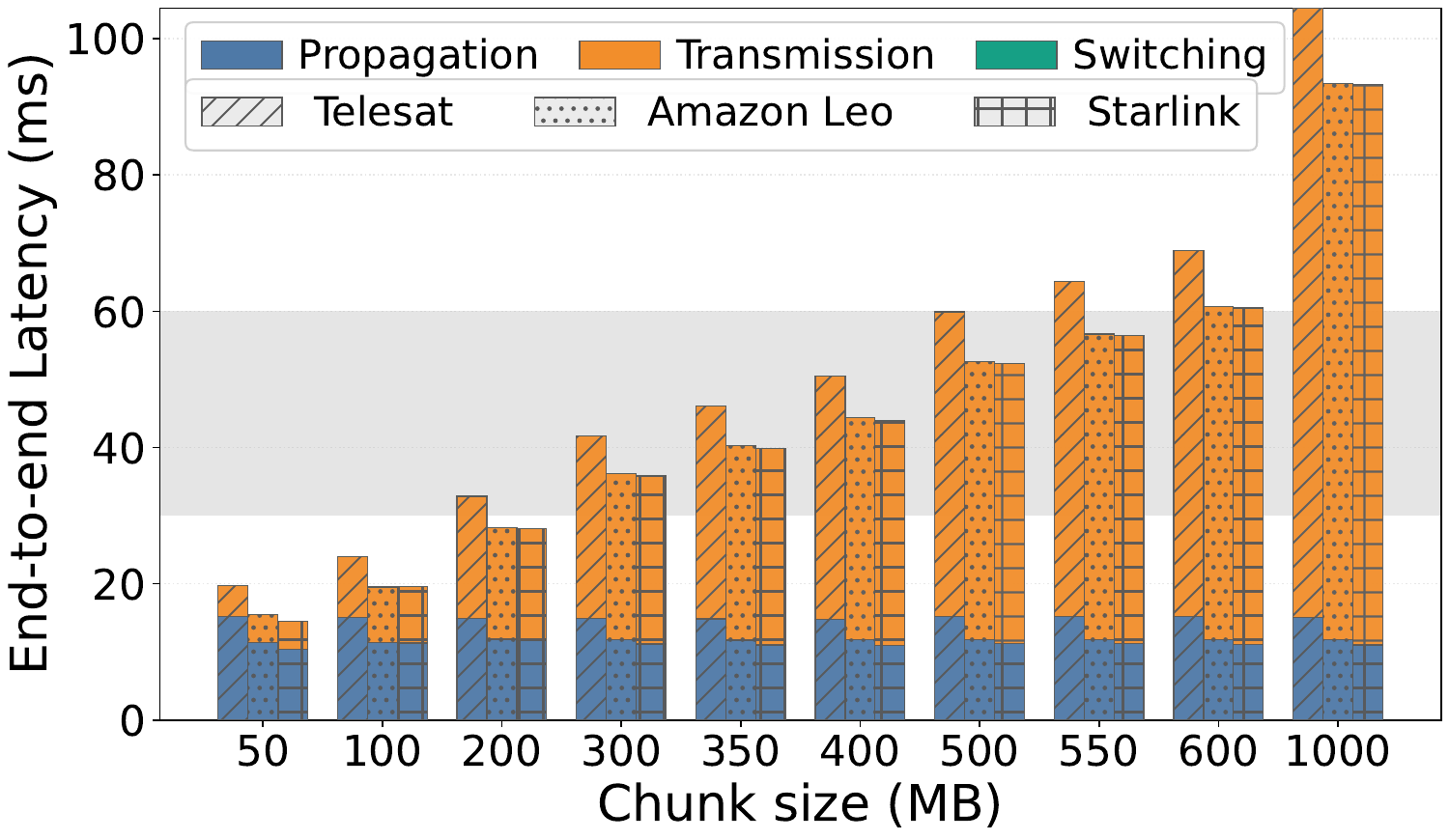}
    \caption{Electro-optic and Inp-based switches}
    \label{fig:fig2.a}
\end{subfigure}
\begin{subfigure}[b]{0.32\textwidth}
    \includegraphics[width=\textwidth]{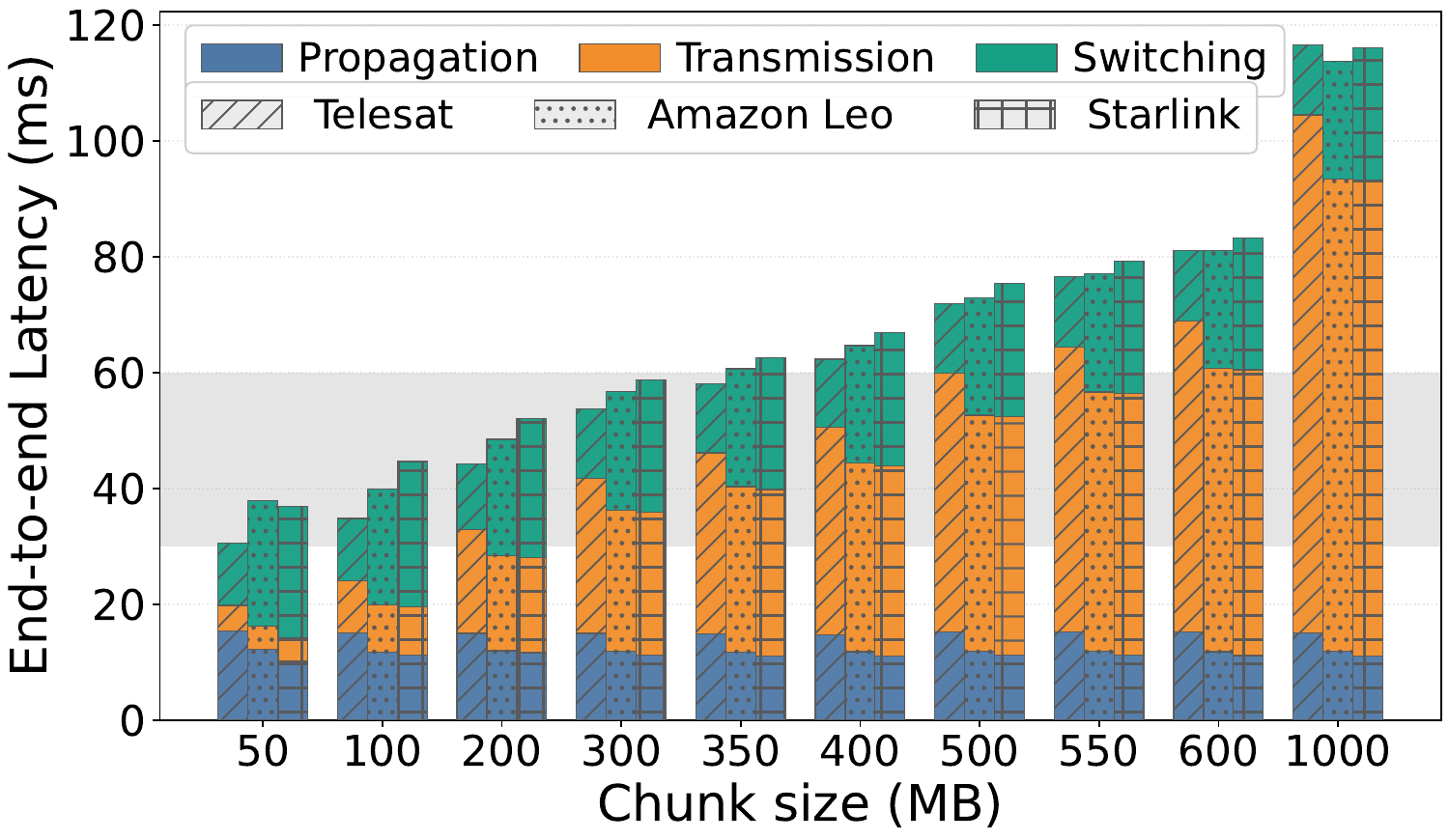}
    \caption{MEMS}
    \label{fig:fig2.b}
\end{subfigure}
\begin{subfigure}[b]{0.32\textwidth}
    \includegraphics[width=\textwidth]{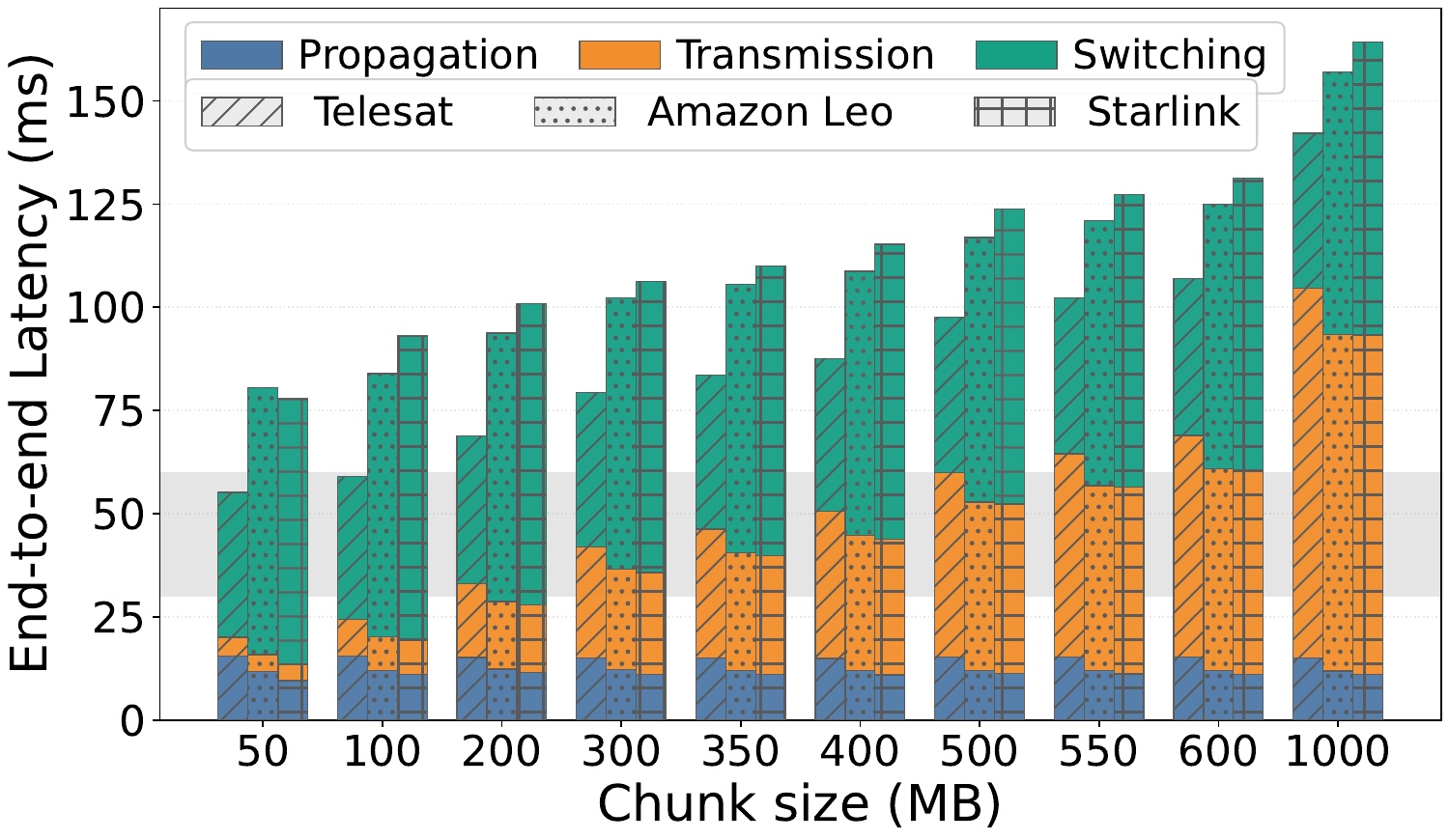}
    \caption{Beam-steering-based switch}
    \label{fig:fig2.c}
\end{subfigure}
\caption{e2e latency breakdown for different chunk sizes,  switching technologies, and constellations.}
\label{fig:fig2}
\end{figure*}

To define a benchmark for a realistic e2e latency threshold in today's constellations, we use the Ookla and Human Settlement Proxy datasets, which show latency measurements that varies roughly between 30 to 60 ms for the five starlink markets with lowest latency in the third quarter of 2025 \cite{dano2026ookla}. We use this range as the latency constraint.
%\hl{DESCIBE WHICH BENCHMARK LATENCY YOU THEN USE AND HOW DO YOU USE THE DATA SETS ABOVE}
Table~\ref{tab:const} shows the constellations studied, with one high-altitude constellation with fewer satellites (Telesat), which can be assumed to offering longer uplinks and downlinks transmission windows, and higher propagation delays. Two denser constellations, i.e., Amazon Leo shell 1 and Starlink phase 1, are also studied. In all constellations, we assume that each satellite node is equipped with four optical terminals for communication with neighboring satellites, arranged in a Grid+ topology, along with an additional port for OGS connectivity. Four OGSs covering Europe are considered \cite{Farley:24}, including Maspalomas (27.76$^\circ$,-15.59$^\circ$), Athens (37.98$^\circ$,23.72$^\circ$), Milan (45.4642$^\circ$, 9.1900$^\circ$), and Bilbao (43.2630$^\circ$, -2.9350$^\circ$).

\begin{table}[h]
\centering
\caption{Constellations configuration.}
\label{tab:const}
\setlength{\tabcolsep}{3.5pt}
\renewcommand{\arraystretch}{1.0}
\begin{tabular}{l |c c c}
    \hline
    \textbf{Constellation} & \textbf{Starlink Phase 1} & \textbf{Telesat} & \textbf{Amazon Leo Shell 1} \\
    \hline
    Satellites          & 1584 & 220  & 784  \\ \hline
    Orbital Planes      & 72   & 20   & 28   \\ \hline
    Inc. ($^\circ$)     & 53.00 & 50.88 & 33.00 \\ \hline
    Alt. (km)           & 550  & 1325 & 590  \\ \hline
    Average hop count   & 2.85  & 1.46 & 2.56  \\
    \hline
\end{tabular}
\end{table}

Event-driven simulations are carried out using OMNeT++ v6.3.0. and the simulation results are reported with a 95\% confidence interval. Traffic chunk arrivals are simulated based on negative Poisson distribution with holding times approximated to match the chunk size transmission. The holding time of each chunk corresponds to the time taken to transmit the chunk across its e2e path. For each simulation, we randomly generate chunks, all with a fixed chunk size $B$, for a source-destination OGS pair. In order to have comparable network loads for different chunk sizes, we normalize the inter-arrival rate proportional to the chunk size $B$. This is done to maintain a similar load over all simulations with different chunk sizes.

%To keep loads comparable across traffic chunk sizes, $\lambda$ is set based on the bottleneck channel capacity and the corresponding transmission delay for each size. 

%\hl{this past para is not related work is merely some tech survey. this needs to be put in the context of the paper or included in the results or somewhere, it does not seem to fit. alternatively, the subsection needs to be renamed into background and related work. I think this shoudl be moved to Results section. also power etc...is there related work??}
Table~\ref{table:switches} shows the parameters of the switching devices studied~\cite{GLSUN-MEMS, Agiltron, Polatis-6000s, Feyisa-soa-2022}. From a technological perspective, beam-steering and MEMS-based switches are mature, but exhibit comparably slower switching times (ms range) and larger physical dimension. Mature electro-optic and experimental SOA integrated photonic (InP) switches enable $\mu s$ and ns-scale reconfiguration and compact integration, making them attractive for size- and weight-constrained satellite nodes.
%Their ultrafast switching capability is particularly advantageous in OBS and fast circuit switching environments, where reduced reconfiguration time directly improves performance.

\begin{table}[h]
\centering
\caption{Selected optical switching fabrics and parameters.}
\label{table:switches}
\setlength{\tabcolsep}{2.5pt}
\renewcommand{\arraystretch}{1.0}
\begin{tabular}{l| c c c c}
\hline
\textbf{Feature} 
& \makecell{\textbf{Beam-steering}}
& \makecell{\textbf{MEMS}}
& \makecell{\textbf{Electro-optic}}
& \makecell{\textbf{InP-based}} \\
\hline

Ref.
& \cite{Polatis-6000s}
& \cite{GLSUN-MEMS}
& \cite{Agiltron}
& \cite{Feyisa-soa-2022} \\

\hline

Speed 
& \SI{25}{\milli\second} 
& \SI{8}{\milli\second} 
& \SI{0.1}{\micro\second} 
& \SI{5.2}{\nano\second} \\

\hline

Power Cons.
& \SI{5}{\watt} 
& \SI{1.25}{\watt} 
& \SI{10}{\watt} 
& \SI{0.58}{\watt} \\

\hline

Insertion Loss
& \SI{1}{\decibel} 
& \SI{2.6}{\decibel} 
& \SI{3.5}{\decibel} 
& \SI{0}{\decibel} \\

\hline

Readiness
& Commercial
& Commercial
& Commercial
& Experimental \\

\hline
\end{tabular}
\end{table}

OMNeT++ v6.3.0 offers the OS3 library which is employed to generate satellite constellations shown in Table~\ref{tab:const}, including their propagation models based on TLE data obtained via~\cite{TLEGen}. 
We implement first-fit as the wavelength assignment policy ($\lambda_1 - \lambda_4$); the channel estimator is used as as described in Sec. \ref{subsec:control}. For routing, we adopt a k-shortest path algorithm ($K=5$), where the physical link length is used as edge weight. The scheduler is assumed to use the offset time $\tau_{off}$ as the time limit for buffering a chunk when no channel is available, and the chunk is dropped when reaching this limit.
%\hl{YOU NEED TO EXPLAIN HERE EXACTELY HOW YOU DO WAVELENGTH ASISMGEN AND SCHEDULING NOT THAT YOU CODIED IT AND "IT WORKS"} 
%Wavelength assignment is performed using a first-fit strategy, selecting the earliest available data wavelength ($\lambda_1 - \lambda_4$). Routing then follows the K-shortest paths algorithm with $k=5$, where edge weights are the physical links lengths, 
%\hl{Traffic is modeled with a negative exponential inter-arrival distribution  which corresponds to a Poisson arrival process with arrival rate $\lambda$, which represent the number of bursts generated per second}. 
A guard time is required between different control messages when transmitted over the same link and control channel $\lambda_C$, and a fixed offset time $\tau_{off}$ is added between the launching of a control message and its corresponding chunk.
The remaining simulation parameters are summarized in Table~\ref{tab:paras}.

\begin{table}[h]
\centering
\caption{Simulation parameters.}
\label{tab:paras}
\setlength{\tabcolsep}{5pt}
\begin{tabular}{l | c}
    \hline
    \textbf{Parameter} & \textbf{Value} \\ 
    \hline
    Central wavelength & 1550 nm (C-band)\\ \hline
    WDM channel spacing & 100 GHz \\ \hline
    % Routing alg. & KSP, $k=5$\\ \hline
    OGS gain & 125 dBi  \\ \hline
    OGS aperture size & 0.8 m \\ \hline
    Satellite power & 0.7 dBW\\ \hline
    Satellite gain & 120 dBi\\ \hline
    Visibility & 60 km (clear sky)  \\ \hline
    Nr. of wavelengths & 5 per port (4 data + 1 control) \\ \hline
    Nr. of ports per node & 4 (ISL) + 1 (OGS), bidirectional \\ \hline
    Guard time & 100~ns\\ \hline
    Offset time & 400~ns\\ \hline
    Processing delay & 10~ns  \\ \hline
    Chunk size & 50 - 1000~MB\\ \hline
   Minimum elevation angle & $30^\circ$ \\
   \hline
\end{tabular}
\end{table}

\begin{table}[th]
\centering
\caption{Evaluation of the maximum feasible burst size across different constellations and switching devices.}
\label{tab:conc}
\setlength{\tabcolsep}{2.5pt}
\renewcommand{\arraystretch}{1.05}
\begin{tabularx}{\columnwidth}{l| X c c c c}
    \hline
    \textbf{Switch type} & \textbf{Const.} & \textbf{Traffic} &
    \textbf{CFR (\%)} & \textbf{E2E L (ms)} & \textbf{EE (Gbits/W)} \\
    \hline
    \textbf{InP} & Telesat & 500MB & 99.02 & 59.91 & 4.55 \\
    & Amazon  & 600MB & 99.32 & 60.68 & 3.22 \\
    & Starlink & 600MB & 99.17 & 60.47 & 2.89 \\
    \hline
    \textbf{Electro-optic} & Telesat & 500MB & 99.02 & 59.91 & 0.26 \\
    & Amazon  & 600MB & 99.32 & 60.68 & 0.18 \\
    & Starlink & 600MB & 99.18 & 60.47 & 0.16 \\
    \hline
    \textbf{MEMS} & Telesat & 350MB & 98.84 & 58.04 & 1.48 \\
    & Amazon  & 350MB & 98.99 & 60.65 & 0.87 \\
    & Starlink & 300MB & 98.67 & 58.70 & 0.66 \\
    \hline
    \textbf{Beam-steering} & Telesat & 100MB & 97.60 & 59.03 & 0.11 \\
    \hline
\end{tabularx}
\end{table}

\par Fig.~\ref{fig:fig1} presents the CRF (Eq.~\ref{eq:BR}) on a logarithmic scale for simulations with different bust sizes $B$, in which $B$ is fixed for each simulation, and varies from 50~MB to 1~GB across different simulations. The beam-steering-based switch yields the lowest CFR in the experiment, with values lower than 98\% for chunk sizes up to 400MB in all constellations due to the slower switching mechanism. %Scenarios with small chunk sizes have more frequent switching due to the reduced inter-arrival rate, and this causes an overhead in devices with lower switching speed.
%which is the dominant factor in the e2e latency shown in Fig.~\ref{fig:fig2.c}. 
The same behavior occurs for the MEMS-based switch, up to a chunk size of 50~MB in Telesat (Fig.~\ref{fig:fig1.a}), and 200~MB in Amazon LEO (Fig. \ref{fig:fig1.b}) and Starlink (Fig. \ref{fig:fig1.c}). This early convergence with Telesat happens due to its broader coverage and the lower number of hops crossed, which reduces the use of switches and consequently the switching delay.
%and making propagation the dominant delay for smaller chunk sizes in Fig.\ref{fig:fig2.b}. 
On the other hand, $\mu s$- and ns-switches show similar CFR values, above 99\% across different chunk sizes, as these devices add negligible switching delay.

\par Fig.~\ref{fig:fig2} illustrates the latency breakdown for different switching fabrics. 
%We consider e2e latency range between 30 and 60 $ms$ as a constraint and target in performance evaluation \cite{dano2026ookla}. 
The Telesat constellation has the highest altitude and the highest propagation delay, followed by Amazon LEO and Starlink, respectively. For Inp- and electro-optic-based switches (Fig.~\ref{fig:fig2.a}), the maximum feasible traffic chunk sizes under the 60ms threshold are 600~MB (Amazon LEO and Starlink, with CFR values of 99.32\% and 99.17\%, respectively) and 500~MB (Telesat, with CRF of 99.02\%). 
%Chunk sizes of 600 MB reach a 99.17\% feasibility in Starlink, and 99.32\% in Amazon Leo. Telesat, in turn, reaches from 99.02\% feasibility with 500~MB chunks.
Therefore, Amazon Leo with 600~MB demonstrates 0.15\% lower feasibility and 0.21~ms higher e2e latency than Starlink. For the ms-scale switches, the adoption of MEMS-based switches (Fig.~\ref{fig:fig2.b}) causes a maximum chunk size of 350 MB for Telesat and Amazon LEO, and 300 MB for Starlink. The increased switching delay reduces the time available for transmission and pushes for smaller chunk sizes.

%The lower burst size for Starlink occurs due to the higher switching required, caused by the higher hop count. 
Finally, beam-steering-based switches (Fig.~\ref{fig:fig2.c}) only reach the e2e latency threshold in Telesat for chunks up to 100 MB size, due to the high added switching delay, while for the other constellations, chunk sizes of 50 MB already exceed the 60~ms e2e delay threshold. For beam-steering switches, the transmission delay only becomes dominant for bigger chunks (1~GB).  Reduced chunk sizes are dominated by switching delay and have often optical channels in an idle state.

We further study the impact of energy efficiency (EE) enabled by the resource allocation solution in the different scenarios, defined as the total number of successfully delivered bits divided by the total power consumption of optical switching hardware, normalized by the chunk losses, i.e.,

\begin{equation}\label{eq:EE}
    \mathrm{EE}~(\text{bits/W}) = \frac{{\sum_{\mathrm{chunks}}}Bits_{\mathrm{received}}}{\sum_{\mathrm{sats}} P_{\mathrm{switch}}} * \mathrm{CFR}
\end{equation}

Based on the results, the maximum feasible chunk sizes for each switch and constellation were selected and summarized in terms of feasibility (CFR),  latency (E2E L), and energy efficiency (EE) in Table~\ref{tab:conc}. The Inp-based switch achieves the best performance in all evaluated metrics. Among the other solutions, MEMS-based switches deployed in Telesat has the highest energy efficiency for 350 MB chunk sizes. 
%Therefore, these results indicate that the appropriate choice of chunk size depends on constellation layout, switching-fabric, and system objectives. Therefore, jointly optimizing chunk size and switching technology provides an opportunity to balance chunk feasibility, latency, and energy efficiency.

\section{Conclusion} \label{sec:conclusion}

We investigated the performance of chunk size variation applied to satellite constellations, based on different scenarios of all-optical switching devices. The results show that ms-scale switches incur lower feasibility for small chunks and require large chunks to meet the 30-60 ms latency.  In contrast, $\mu$s- and ns-switches achieve a feasible chunk size of 500MB for Telesat and 600 MB for Amazon Leo and Starlink, while maintaining the 60~ms latency threshold and chunk feasibility ratio $\textgreater ~99 \%$.
% Considering energy efficiency, ms-scale switches outperform certain fast-switching alternatives in some cases, and the ultimate choice of data chunk size must consider a joint evaluation of feasibility ratio, e2e expected latency, and energy efficiency. Future research points to the evaluation of extra traction added to e2e latency by other devices, such as wavelength converters and elastic transceivers, and the design of models for optimizing routing, wavelength assignment, burst sizing, and scheduling in the investigated architecture.

\section*{Acknowledgment}

This work was partly funded by the European Space Agency (ESA) in the framework of the project NEXON (activity number 1000042074).

\bibliographystyle{IEEEtran}
%\bibliography{./bibliography/IEEEabrv,mybib}
\bibliography{mybib}

\end{document}